\documentclass[english,aps]{revtex4}
\usepackage{graphicx}
\usepackage{amssymb}

\makeatletter

\usepackage{babel}
\makeatother
\begin{document}

\preprint{FQXi Essay Contest - The Nature of Time - Fourth Classified}

\title{\LARGE  The Nature of Time: from a Timeless Hamiltonian Framework to Clock Time of Metrology}

\author{\large Enrico Prati}

\affiliation{Laboratorio Nazionale Materiali e Dispositivi per la Microelettronica,
Consiglio Nazionale delle Ricerche - Istituto Nazionale per la Fisica della Materia, Via Olivetti 2, I-20041
Agrate Brianza, Italy}

\begin{abstract}
\large
The problem of the Nature of Time is twofold: whether or not time is a fundamental quantity of Nature, and how does clock time of metrology emerge in the experimental description of dynamics.
This work strongly supports the fundamental timelessness of Nature. However, the correct view that physics is described by relations between variables does not address the second problem of how time does emerge at the macroscopic scale on the ground of a timeless framework. 
In this work ordinary Hamiltonian dynamics is recast in a timeless formalism capable to provide a definition of parameter time on the basis of the only generalized coordinates, together with the Hamiltonian invariance on trajectories, and a variational principle.  Next, by relaxing the assumption of periodicity of real clocks to the only cyclicity in the phase space, the second problem is addressed. Physical systems, if complex enough, can be separated in a subsystem whose dynamics is described, and another cyclic subsystem which behaves as a clock. The dynamics of the first is mapped in the states of the second cyclic subsystem which provides a discrete approximation of the parameter time. Useful clocks fulfill a stability prescription which guarantees that dynamics is expressed by simple laws also in terms of metric time as naturally happens for parameter time. The two ideas provide a unitary framework capable to account the fundamental timelessness of Nature, and the experimental evidence of time evolution in macroscopic systems experienced by the observers.
\end{abstract}

\maketitle

\large
\section{\large Introduction}
\label{intro}

The odd nature of time arises from a large number of reasons, first of all the discrepancy between the satisfactory description of dynamics in terms of time evolution, and the fundamental timelessness of general relativity and canonical quantum gravity \cite{DeWitt67,Elze04,Rovelli94II,Rovelli06,Isham06,Anderson07,Albrecht07}. 
Several contradictions emerge when different theories are compared.
Classical and quantum mechanics
use time as an external parameter to generate a strongly continuous
unitary group of transformations $U(1)$ \cite{ReedSimon72}. Some
theories like the special and the general relativity, quantum field
theory, and canonical quantum gravity, use time as the negative metric
signature coordinate of a 4-dimensional differential manifold. It
is the coordinate associated to a smooth foliation of codimension
1 called topological time \cite{Itin04}, to which observables and
operators are associated according to locality. Time coordinate 
is subject with space coordinate to Lorentz or
more general transformations \cite{Rovelli06}. 
In metrology, clock time is a metric
time operatively defined at rest. In statistical physics,
time is associated to the concept of irreversibility and its flow
to the entropy of a closed system \cite{Lebowitz93,Lebowitz93II, Rovelli93}.
Such different views are mainly due to implicit assumptions of properties that the system under investigation is supposed to own, such as time reversal symmetry, covariance, locality, monotonical growth of entropy and others.

When considering the problem of the nature of time, there are implicitly two distinct problems to address. The first (i) it is about whether or not time must be included in the list of fundamental quantities of Nature, well defined at all the possible energy and length scales. The second is (ii) the description of the emergence of time metrology based on operatively defined clocks, and the explanation of time evolution experienced by an observer.
The present work strongly supports the view that time is not a fundamental quantity of Nature. I will show that Hamiltonian mechanics, which governs the dynamics of generalized coordinates or quantum fields, is rigorously well defined without the concept of time. As a consequence, it becomes apparently even more difficult to solve the problem (ii), basically because time disappears from the list of observable quantities, but there is a field of metrology entirely devoted to time and frequency measurements (capable to quantify the common experience of time flow).

A good theoretical model capable to be predictive and satisfactory without time answers only partly to the problem of the Nature of Time: it says what time is not \cite{Rovelli06}. In the present work the answer to (i) is addressed, with a particular attention in the definition of the time parameter in a Hamiltonian system in terms of other quantities. Next, also the second part (ii) of the problem is addressed, in order to account the experimentally measured and experienced clock time. The connection between the experimental clock time and the theoretical parameter time is the main goal of the present contribute. There is a recurrent misleading identification between clock time as measured by time and frequency references in a laboratory, and the parameter with respect to which entropy and disorder grow. The observation that thermodynamics leads towards equilibrium states has nothing to do with the concept that motion of its constituents can be described as a function of a clock time. The latter is operatively defined by the prescriptions for realizing the clock itself.
Therefore, clock time requires to be investigated by looking at the general properties of suitable subsystems acting as clocks according to their properties instead of thermodynamic properties of statistical states.

The fact that time is not a measurable quantity can be clarified as follows \cite{DeWitt67,Gerlach69,Banks85,Halliwell85,Brout89,Elze04}. One observes that a 
clock measures with some uncertainty an hypothetical 'true' external
time $t$ as a classical quantity $T_{i}(t)$ where the index $i$ spans the clocks. The other observable
quantities $O_{j}(t)$ are detected as $O_{j}(T_{1})$, $O_{j}(T_{2})$,.... where $j$ spans the observables.
However, the clock used to label the dynamical quantities of the system is in turn object of a measurement which establishes its value, accuracy and stability, by means of another clock. Consequently, being the second clock subject to the same check by the first, a two-clock time measurement is required to determine the fractional
frequency stability from the Allan variance of both and a reference standard
is expressed by $T_{2}(T_{1})$ and  $T_{1}(T_{2})$,\cite{Rovelli06,Levine99}
without any explicit use of $t$. In other words, a clock is not capable to measure such hypothetical external parameter time, but only self consistent quantities assisted by the recursive definition of $period$.

In the following we addresses the problem of explaining the macroscopic correspondence of clock time with parameter time of dynamics and we provide a universal definition of time for a Hamiltonian system in terms of generalized coordinates change in the phase space.

I will concentrate on a variational
approach which enables the introduction of time in a physical theory in two steps. The method provides a parameter, called parameter time, which does not correspond to a specific observable quantity. However it can be put in correspondence with measurable quantities via cyclic phenomena. This is achieved by dividing a system in opportune subsystems. The present approach partially recalls the distinction presented in Ref.\cite{Elze04} between parametric (proper) time and discrete physical time. Differently from there, here no compactified extra-dimensions are required to introduce a detector operator, neither a lapse function or other parameters to appear in the Lagrange function.

In Section II the definition of parameter time in the framework
of timeless Hamiltonian theories is presented. Time emerges as the
natural parameter after one imposes a variational principle on a timeless
action. The approach is applied in the subsection 2.1 to classical mechanics,
and extended to quantum field theory in the subsection 2.2. Section III is devoted to connect the parametric time with clock metric time measured by means of realistic devices.
In Section IV the results are compared with other approaches and discussed.

\section{\large Parameter Time in a Hamiltonian Timeless Scenario}
\label{sec:1}

The Maupertuis(-Euler-Lagrange-Jacobi) \cite{Arnold,Landau} action principle generates the dynamics without explicitly
using time (for a review on the Maupertuis principle, see
\cite{Gray99}) in the Hamilton theory. Furthermore, we consider only parameter independent Hamiltonians, consistently with the hypothesis of dealing with a closed system.
Such assumptions allow to express the variational principle, the Hamiltonian and the generalized coordinates in a timeless framework. The imposition of both the variational principle and the stationarity of the Hamiltonian individuate a special parametrization among all the possible parametrizations. Such parametrization is useful to describe dynamics. In the following the corresponding parameter is indicated by $\sigma$ and corresponds to the parameter $\tau$ of Ref. \cite{Arnold}, and to parameter time $t$ in ordinary Hamiltonian theory. The main difference from the latter is given by its derivation in a timeless framework. The capability of defining Hamiltonian mechanics without the concept of time will require consequently that some extra hypothesis are assumed in order to provide a definition of clock time. Its correspondence with the parameter $\sigma$ is defined and discussed in the next section.

\subsection{\large Parameter time in classical mechanics}
\label{sec:2}

To derive parameter time from the Maupertuis principle, we have to recast in the timeless framework the ordinary derivation of Hamilton equations of motion from a variational principle on asynchronous varied trajectories.
Contrarily from Lagrangian formalism, expressed by $n$ independent coordinates $q_{i}$ and their time derivative $\dot{q_{i}}$, the Hamiltonian approach allows to set first
order differential equations by virtue of the independence of the
momenta $p_{i}$ from the coordinates $q_{i}$. The number of independent
coordinates raises to $2n$. 
The time independent Hamiltonian $H(\mathbf{p},\mathbf{q})$ is a function
of the generalized three dimensional coordinates $\mathbf{p}$ and
$\mathbf{q}$. The independence of $H$ from $t$ reduces the degrees of freedom to $2n-1$. It is necessary to assume that it exists a set of trajectories
in the coordinates space $\mu$ for which $H$ is constant. 

A generic parametrization of the points
of the trajectories is set by a label $\lambda$ so that $q_{i}=q_{i}(\lambda)$
and $p_{i}=p_{i}(\lambda)$ where all such functions belong to $C^{2}$
on the interval $[\lambda_{A},\lambda_{B}]$$\in\mathbb{R}$. The
Hamiltonian $H(\mathbf{p},\mathbf{q})$ does not depend explicitly
on $\lambda$. In order to impose a variational principle
on the trajectory it is now considered a variation that is normally
used to impose asynchronous varied trajectories in canonical formalism
to derive Hamilton equation from the Maupertuis principle.
A new parametrization $\sigma$ of
the generalized coordinates and of $\lambda$ is
now defined, under the condition that $\frac{d\lambda}{d\sigma}\neq0$
on $[\sigma_{A},\sigma_{B}]$. 

The parametrization $\lambda$ is generic and arbitrary under the given assumptions, necessary only as a starting point. On the contrary the parametrization $\sigma$ is strictly related to the stationarity of the Hamiltonian.
Such distinction is therefore fundamental in the present derivation and it represents a subtle principle and technical difference from the approach of Ref.\cite{Barbour94, Barbour94II}

The stationarity of the action is imposed:

\begin{equation}
A=\int p_{i}dq_{i}
\end{equation}

where the Einstein summation on the repeated indexes is adopted and $i=1,2,3$. The Maupertuis variational principle reads 

\begin{equation}
\delta A=\delta\int p_{i}dq_{i}=0
\end{equation}

The imposition of the stationarity of the action is given by the variation
of the trajectories. Neglecting as usual second order perturbations  and integrating by parts where necessary, one
has:

\begin{equation}
d\sigma=\left(\frac{\partial H}{\partial p_{i}}\right)^{-1}dq_{i}=-\left(\frac{\partial H}{\partial q_{i}}\right)^{-1}dp_{i}
\end{equation}

under the hypothesis that $\left(\frac{\partial H}{\partial p_{i}}\right) \ne 0$ and $\left(\frac{\partial H}{\partial q_{i}}\right) \ne 0$.
They differ from the Hamilton equations since $\sigma$ does not represent the  macroscopic metric time. On the contrary, it only represents the natural parameterization of the system imposed by the energy conservation.

We now briefly discuss the results before extending to quantum field theory.

First, there is one natural parameterization ($\sigma$) which can be
defined by virtue of the properties of the Hamiltonian along the trajectory,
common among all the conjugated coordinate pairs $(p_{i},q^{i})$. Since
all the information about the trajectory and the Hamiltonian are assumed
to be known, the only free parameter of such equation is the label
$\sigma$ which can be identified with local parameter time of dynamics. Notice that
such parameter is not the observable quantity measured by clocks.
Second, the parameter $\sigma$ provides the measurement of the change of
the system along the trajectory. The ratio between the amount of change
of the conjugate variables $dq_{i}$ and $dp_{i}$ during motion is
weighted by the ratio between $\left(\frac{\partial H}{\partial p_{i}}\right)^{-1}$
and $-\left(\frac{\partial H}{\partial q^{i}}\right)^{-1}$. The quantity
$d\sigma$ measures the amount of change along the two generalized
coordinates when energy conservation holds.

\subsection{\large Parameter time emerging in Quantum Field Theory }
\label{sec:3}
The most convenient formalism to extend the action principle to general relativity and to quantum mechanics is the extended presymplectic approach \cite{Rovelli06}. There, dynamics is expressed on the unparameterized curve $\gamma$ in the relativistic configuration space $C=\mathbb{R}\times C_0$, where $C_0$ is the $m$-dimensional space of coordinates $q^i$, which extremizes the integral 

\begin{equation}
A[\gamma]=\int_{\gamma}\theta
\end{equation}

where 

\begin{equation}
\theta=p_i dq^i + p_t dt
\end{equation}

is the natural one-form defined on the cotangent space $T$*$C$ and the constraint 

\begin{equation}
H(q^i,t,p_i,p_t)=0
\end{equation}

where $H$ is the relativistic Hamiltonian. In the extended presymplectic formalism, the variational principle reads:

\begin{equation}
\delta A[\gamma]=\delta \int_{\gamma}\theta=0
\end{equation}

Such principle allows a quantum extension, which goes beyond the scopes of the present section. 
Both the lagrangian and the extended presymplectic formalism consider time as a part of the manifold where physics is defined. Time $t$ or $x_0$ assumes a role comparable to that of space, even when starting with an unparameterized curve as happens in presymplectic approach. Technically, since the action admits invariance under reparameterization of time (spacetime in relativistic domain), it does not represent a problem. 
Here, in order to avoid the use of the concept of time, the configuration space is only $C_0$ instead of $C=\mathbb{R}\times C_0$ and the extended configuration space will only include fields and their conjugate momenta (generalized fields).

The quantum field theory is generally given in terms of (anti)commutation
relations. The physical content of a theory is well expressed under
the manifestly covariant Lagrangian formalism, but the physics can 
be totally described in terms of S-matrix formalism, after one has
imported the physical content in the Hamiltonian approach. Here, in
order to ensure the continuity with the previous analysis, quantum
field theory is considered in Hamiltonian formalism. A Hamiltonian
operator $H=\int d^{3}x\mathcal{H}$ is given, where $\mathcal{H}$
is the Hamiltonian density. The Hamiltonian operator $H$ acts as
a constraint for quantum field dynamics. In quantum field theory,
such a constraint corresponds to being on the mass shell.
The action, in terms of a
quantum fields $\psi_{i}(x)$ and the conjugate coordinates $\pi_{i}(x)$,
can be re-expressed as: 

\begin{equation}
A=\int d^{3}x\int d\psi_{i}\pi_{i}
\end{equation}

where the Einstein summation on the repeated indexes is adopted. The
roman index spans on the space dimensions 1, 2, and 3. To define time
as the natural parameterization of change in the generalized coordinate space $\mu_{Q}$, the points of the trajectories $f(q_{i},p_{i})=0$ are
replaced in QFT by space configurations of the generalized field $Q=\left(\psi_{i}(\mathbf{x}),\pi_{i}(\mathbf{x})\right)$
in $\mu_{Q}$ . In the classical case neighboring
position and momentum states are associated to the parameter $\sigma$,
while in QFT $\sigma$ labels the generalized field with support in $\mathcal{\mathbb{R}}^{3}$. Two arrays of fields variate the quantum fields and their conjugate fields respectively.
As in the previous case, the extremality of the action is obtained under the condition that:

\begin{equation}
d\sigma=\left(\frac{\delta\mathcal{H}}{\delta\pi_{i}}\right)_{\psi_{i}}^{-1}d\psi_{i}(\mathbf{x})=-\left(\frac{\delta\mathcal{H}}{\delta\psi_{i}}\right)_{\pi_{i}}^{-1}d\pi_{i}(\mathbf{x})
\end{equation}

Parameter time can be defined as the rate of change of the fields $\psi_{i}$
and their conjugate variables $\pi_{i}$ via the factor $\left(\frac{\delta\mathcal{H}}{\delta\pi_{i}}\right)_{\psi_{i}}^{-1}$and
$-\left(\frac{\delta\mathcal{H}}{\delta\psi_{i}}\right)_{\pi_{i}}^{-1}$
respectively. The parameter $\sigma$ belongs to $\mathbb{R}$ by
construction. The parameterization of the field distribution is locally
achieved by tagging neighboring configurations with the parameter
$\sigma$. Parameter time is therefore built in analogy with the
classical case even in the microscopic limit when small variations
of the fields are considered. Such a construction is compatible with
all the canonical Hamiltonian theories and naturally provides parameter
time in timeless models. An important remark is that parameter time is 
by construction defined for those particles which are on the mass shell.

\section{\large Clock time}
\label{sec:4}
$\sigma$ has the property of providing a privileged parameterization suitable for describing dynamics, but it is not an observable quantity. In order to explain the macroscopic experience of time in complex systems, an observable quantity $T$ is built. $T$ realizes an experimentally measurable discrete approximation of $\sigma$.
Since (metric) time is operatively defined by clock standards based on the period of an oscillator, it is only defined in such systems complex enough to contain a subsystem acting as such a clock.
However, the concept of $period$ has to be relaxed to the concept of $cycle$ in the phase space $\mu$ or in $\mu_{Q}$.
Indeed, the definition of periodicity implicitly assumes that an external time is available in order to compare a period with the next one, which is meaningless in a timeless framework.
Defining the $clock$ $time$ $T$, measured for example by atomic clocks, corresponds to label simultaneous occurences in the phase space of two or more subsystems where one is identified as the clock. The clock corresponds to the cyclic subsystem, as defined below. The macroscopic time measured by a subsystem is a function of the subsystem itself. It is a matter of the experimentalist to choose suitable cyclic subsystems (macroscopic clocks) in order to provide a good approximation of the parameter time $\sigma$. The dynamics of the $i$-th observable $O_i$ will consequently be expressed by the simple law involving $\sigma$:

\begin{equation}
O_i (T) \cong O_i (\sigma)
\end{equation}

Let's consider a Hamiltonian system $S$ separable in two independent
subsystems $S_{1}$ and $S_{2}$, so that all the states are represented
by factorized (eigen)states of their respective Hamiltonian $\psi_{1}\otimes\psi_{2} \in  H_{1}\otimes H_{2}$ where $H_1$ and $H_2$ are the Hilbert spaces of the subsystems 1 and 2 respectively.
From the previous analysis, the system $S$ owns a unique natural parameter
time $\sigma$ which is well defined also separately for the two subsystems by construction.
We now define the properties required by the system $S_1$ to act as a clock in $S$ in order to describe dynamics in $S_2$.
For a given $\bar{\sigma}$, a state $\psi \in H_{1}\otimes H_{2}$ consists of the tensor product of the state $\psi_{1}(\bar{\sigma}) \in H_1$ and the state $\psi_{2}(\bar{\sigma}) \in H_2$. 
We say that $\bar{\psi_{1}}$ has multiplicity $\kappa_{AB}$ on the interval $(\sigma_A,\sigma_B)$ if there are $\kappa_{AB}$ values of $\tilde{\sigma}_i \in (\sigma_A,\sigma_B)$ such that $\psi_{1}(\tilde{\sigma}_i)=\bar{\psi_{1}}$ where $i \in (0,\kappa_{AB})$. 
We say that the subsystem $S_1$ is cyclic in the phase space if

\begin{enumerate}
	\item {its path in the phase space is closed, } 
	\item {its velocity $|dQ/d\sigma |\neq 0$ and it is smooth, } 
	\item {the multiplicity $\kappa_{AB}$ of a state vector in the System 1 monotonically grows with the interval $(\sigma_A,\sigma_B)$ and it tends to infinity when $\sigma_A \rightarrow - \infty \wedge \sigma_B\rightarrow + \infty$.}
\end{enumerate}

The second requirement grants that the realizations of two contiguous states occur along the $\sigma$
axis by respecting the order of the parameter $\sigma$. The third requirement that the clock never stops and its velocity in the phase space is enough to grant that the number of cycles is not finite.

Given the interval $(\sigma_A,\sigma_B)$, it is now defined the set $\Omega(\sigma_A,\sigma_B) \subset H_2$:

\begin{equation}
\Omega(\sigma_A,\sigma_B)=\left\{\psi_2 (\sigma) \in H_2 | \sigma \in \left(\sigma_A,\sigma_B\right) \right\}
\end{equation}

An arbitrary origin $\sigma_0$ is fixed for the parameter time. We associate to such origin the arbitrary initial states $\bar{\psi}_1 = \psi_1 (\sigma_0)$ and $\bar{\psi}_2 = \psi_2 (\sigma_0)$.
Macroscopic time duration $T^{(S_1)}$ of the interval  $(\sigma_A,\sigma_B)$ measured by the cyclic subsystem $S_1$ is given by the number $k_{AB}$ of states $\psi_2(\sigma) \in \Omega$ so that $\psi_1(\sigma)=\bar{\psi}_1$. More explicitly, one has 

\begin{equation}
T_{AB}^{(S_1)} \equiv k_{AB}
\end{equation}

A good clock has the property of being $stable$ (small standard deviation)  and $accurate$ (high Q factor of the resonance associated to the clock) \cite{Levine99,Heavner05}. Since the accuracy refers to the arbitrary resonance frequency of the time standard (for example the Cesium resonance frequency), the present analysis considers only the requirement of stability. 
Given a target standard deviation $\Sigma$ required in an experiment performed on the subsystem $S_2$ in the interval $(\sigma_A,\sigma_B)$, for an integration time  $\tau$, the clock has to fulfill the following prescription:

\begin{equation}
\epsilon \equiv E^2\left[T_{i,i+1}^{(S_1)}\right]< \Sigma 
\end{equation}

where $E^2$ is the standard deviation and

\begin{equation}
\sigma_{i+1}=\sigma_{i}+\tau
\end{equation}

where $i=0...N_{AB}$ with $N_{AB}=(\sigma_B-\sigma_A)/\tau$.
The definition of clock metric time loses of validity for time intervals $T^{(S_1)}$
comparable with the clock period, and for shorter time intervals. 
Under such hyphotesis, dynamics of observables in the interval     $(\sigma_A,\sigma_B)$ is approximated by the discrete valued equations:
\begin{equation}
x_{\rho} (T_i) \cong  x_{\rho} (\sigma_i \pm \epsilon)=x_{\rho} (\sigma_i ) \pm O_{x\rho}[\sigma_i,\epsilon] 
\end{equation}
\begin{equation}
p_{\rho} (T_i) \cong  p_{\rho} (\sigma_i \pm \epsilon)=p_{\rho} (\sigma_i ) \pm O_{p\rho}[\sigma_i,\epsilon]
\end{equation}

where $\rho=1,2,3$, and $O_{x\rho}[\sigma,\epsilon]$ and $O_{p\rho}[\sigma,\epsilon]$ are higher order quantities in $\epsilon$. Such equations provide the bridge between parameter time of Hamiltonian timeless formalism, and the experimentally defined clock time experienced by observers.

\section{\large Discussion}
\label{sec:5}

Two consequences of the present interpretation of the clock role in the description of the evolution of another subsystem are briefly discussed.
The first consequence deals with the unavoidable semiclassicality of the measurement of a quantum system. Since clock time is by definition fundamentally discrete and it depends on the specific fabrication of the clock, a (macroscopic) measurement of time below one cycle (period) of the time standard is meaningless. At the present time the most advanced available clock technology is given by single ion atomic clocks based on $Al^{+}/Hg^{+}$ with a fractional uncertainty of about $1-2\times10^{17}$ \cite{Itano06}. Adopting such view it implies for example that Planck time scale is an extrapolation, an extention of the concept of clock time beyond its field of definition.
Following the terminology of Kofler and Brukner \cite{Kofler07}, macrorealism (property of a system of being in one or more macroscopically distinct states) and classical (or semiclassical) laws emerge out of quantum physics under the restriction of coarse-grained measurements. The description of time evolution of a system is necessarily semiclassic because the observer is tracking time with a macroscopic system whose fluctuations dominate on the short time scale. Indeed, $T$ is expected to fail as a good approximation of $\sigma$ in the fast decoherence process which occurs during a measurement. 

The second point deals with the clock ambiguity problem, where clock is treated as a subsystem \cite{Albrecht07} like in the present approach. Though, an important distinction connected to the role of the parameter time is done. There, gauge invariance transforms one parameterization into another, so they are all equivalent. This implies that a complex system can be separated in many ways in a part which constitutes the clock, and the rest. Such property reveals the assumption that parameter time and clock time are considered to coincide. Such approach assumes consequently that parameter time is an observable quantity, contrarily to the argument presented in the Introduction on the non-observability of parameter time. Furthermore, gauge invariance, in the most general case, treats spacetime as a whole, while we maintained the two conceptually distinct in our study. Causality emerges only when the stationarity of energy is imposed, so time becomes part of 4-manifold spacetime \cite{Minguzzi06}.

\section{\large Conclusion }
\label{sec:6}

The problem of the Nature of Time consists of two parts: whether or not time is a fundamental quantity of Nature, and how clock time does emerge in the laboratory measurement in spite of a (timeless) theoretical and conceptual framework according to which parameter time is not observable. This work addresses both the two issues by providing an explicit Hamiltonian framework, entirely developed without the concept of time, and by defining cyclic subsystems capable to account the (discrete) definition of clock time used in time and frequency metrology.
By restricting
the attention to closed systems the Hamiltonian is time independent
and the action principle can be expressed in terms of only the conjugate
variables (Maupertuis action principle) without the concept
of time. The assumption of being on the mass shell, or equivalently that the stationarity of energy holds, along the trajectory in the phase space, provides a parameterization which gives the ratio of change 
of conjugate variables
(generalized coordinates $q_{i}$ and $p_{i}$ or generalized quantum
fields $\psi_{i}$ and $\pi_{i}$). Since all the observables are
expressed in terms of such variables, $\sigma$ parameterizes the whole algebra of observables. In order to well approximate with an observable quantity the parameter time $\sigma$ for which the description of dynamics is simple, we introduce the clock time $T$.
Clock time (also called physical time) measured by macroscopic clocks is a coarse grained discrete
quantity  which can be defined in a system $S$ complex enough to contain a subsystem $S_1$ cyclic in the phase space.
The cyclic subsystem acts as a clock reference used for the operative definition of time. Such metric clock time consists of a discrete approximation of the parameter $\sigma$. The stability is a function of the cyclic subsystem adopted to be the clock. The wanted stability is reached by adopting prescriptions in terms of standard deviation on a given integration time. 
In particular, if one considers a system containing
a subsystem which corresponds to a cyclic phenomenon, the configurations
of the rest (the subsystem $S_2$) can be put in correspondence with complete contiguous cycles on the orbit
of the cyclic phenomenon in its phase space. 
$T$ is therefore suitable to quantify the change
of the other observables defined in $S$, by virtue of its capability of discretize $\sigma$ at the desired precision, according to the characteristic of the experimental apparatus. 
To conclude, the present work provides a unitary framework capable to account the timelessness of Nature at a fundamental level, and to explain how clock time can be defined in metrology and experiments, consistently with the dynamics of relations between variables and parameter time evolution itself.

\begin{acknowledgements}
I gratefully acknowledge E. Minguzzi, Y. Itin, P. Zenczykowski, and H. T. Elze for the helpful suggestions
to improve the first and the second version of the manuscript, and to J. Barbour for the useful criticisms.
\end{acknowledgements}

\bibliographystyle{spphys}       


\end{document}